\documentclass[11pt,a4paper]{article}
\usepackage{jheppub}
\usepackage{indentfirst}
\usepackage{amsmath}
\usepackage{bigstrut}
\usepackage{amssymb}
\usepackage{array}
\usepackage{multirow}
\usepackage{floatrow}
\usepackage{etoolbox}
\usepackage{graphicx}
\usepackage{hyperref}
\usepackage{amsmath}
\usepackage{amsthm}
\usepackage{slashed}
\usepackage{array}
\usepackage{epstopdf}
\usepackage{graphicx}
\usepackage{fancyhdr}
\usepackage{float}
\usepackage{subcaption}
\usepackage[dvipsnames]{xcolor}
\usepackage{hyperref}
\usepackage[utf8]{inputenc}
\usepackage[T1]{fontenc}
\usepackage{lmodern}
\usepackage{textcomp}
\usepackage{microtype}

\usepackage{mathtools}
\usepackage{physics}
\usepackage{units}
\usepackage{siunitx}
\usepackage{ulem}

\def\figureautorefname~#1\null{Fig.\,#1\null}

\def\equationautorefname~#1\null{Eq.\,(#1)\null}

\numberwithin{equation}{section}

\newcommand{\clam}{\gamma_c/\gamma_\lambda}
\newcommand{\glam}{\gamma_\lambda}
\newcommand{\gc}{\gamma_c}
\newcommand{\re}{\operatorname{Re}}
\newcommand{\im}{\operatorname{Im}}

\newcommand{\A}{\mathcal{A}}
\newcommand{\M}{\mathcal{M}}

\newcommand{\eq}[1]{Eq.~(\ref{eq:#1})}
\newcommand{\secref}[1]{Sec.~\ref{sec:#1}}

\title{A Note on the Analytic Structure of Celestial Amplitudes}
\author[a]{Jiayin Gu,}
\affiliation[a]{Department of Physics and Center for Field Theory and Particle Physics, Fudan University, Shanghai 200438, China}
\affiliation[a]{Key Laboratory of Nuclear Physics and Ion-beam Application (MOE), Fudan University, Shanghai 200433, China}
\emailAdd{jiayin\_gu@fudan.edu.cn}
\author[b]{Ying-Ying Li,}
\affiliation[b]{Theoretical Physics Department, Fermi National Accelerator Laboratory, PO Box 500, Batavia, IL 60510, U.S.A.}
\emailAdd{yingying@fnal.gov}
\author[c]{Lian-Tao Wang}
\affiliation[c]{Department of Physics and Enrico Fermi Institute, University of Chicago, Chicago, IL 60637, USA}
\affiliation[c]{Kavli Institute for Cosmological Physics, University of Chicago, Chicago, IL 60637, USA}
\emailAdd{liantaow@uchicago.edu}

\date{\today}

\begin{document}
\preprint{FERMILAB-PUB-22-414-T}
\abstract{
Celestial amplitudes, obtained by applying Mellin transform and analytic continuation on ``ordinary'' amplitudes, have interesting properties which may provide useful insights on the underlying theory. 
Their analytic structures are thus of great interest and need to be better understood.  
In this paper, we critically examine the analytic structure of celestial amplitudes in a massless low-energy effective field theory.  
We find that, fixed-order loop contributions, which generate multipoles on the negative $\beta$-plane, in general do not provide an accurate description of the analytic structure of celestial amplitudes.  
By resumming over the leading logarithmic contributions using renormalization group equations (RGEs), we observe much richer analytic structures, which generally contain branch cuts. 
It is also possible to generate multipoles or shifted single poles if the RGEs satisfy certain relations. Including sub-leading logarithmic contributions is expected to introduce additional corrections to the picture. However, without a new approach,  it is difficult to make a general statement since the analytic form of the Mellin transform is challenging to obtain.  
}
\maketitle
\unitlength = 1mm
\section{Introduction}
Celestial amplitude, reinterpreting scattering amplitudes as correlators in the two dimensional conformal field theory (CFT) \cite{Pasterski:2016qvg, Pasterski:2017kqt, Pasterski:2017ylz}, has been extensively studied from various viewpoints, see {\it e.g.} Refs. \cite{Raclariu:2021zjz, Pasterski:2021rjz} for overviews of this rapidly growing field.
Instead of the usual energy eigenstates, celestial amplitude considers the scattering of boost eigenstates which have both UV and IR physics involved \cite{Arkani-Hamed:2020gyp}. With this feature, celestial amplitude violates the basic Wilsonian decoupling intuition and might provide a new route to probe physics in the UV. Celestial amplitude  also has interesting properties such as providing the correspondence between soft theorem of gauge theory and Ward identities in the celestial CFT \cite{Lysov:2014csa, Cheung:2016iub, Donnay:2018neh, Fan:2019emx, 2019-cst, Pate:2019lpp, Adamo:2019ipt, Nandan:2019jas}, characterizing the infinite number of non-trivial symmetries of 4-dimensional gauge and gravitational theories in asymptotically flat spacetime \cite{Guevara:2021abz} that may bring in new perspectives in understanding flat-space quantum field theory and the construction of consistent $\mathcal{S}$-matrix.  Recently in Ref.~\cite{Arkani-Hamed:2020gyp}, celestial amplitudes were applied to a general effective field theory (EFT) with Wilsonian cutoff, and their properties have been studied. Furthermore, with real boost weight $\beta$, a specific dispersion relation for the celestial amplitude was established in Ref.~\cite{Chang:2021wvv}, relating the imaginary parts of the celestial amplitudes to their residues at negative even interger values of $\beta$. Great efforts are still required to fully understand properties of celestial amplitudes, especially how these properties manifest fundamental priciples of quantum field theory.

The celestial amplitude of a certain process for massless external particles can be obtained by taking the usual amplitude (with energy eigenstates) and performing a Mellin transformation on its energy \cite{Pasterski:2017ylz, deBoer:2003vf, Cheung:2016iub}.  More specifically, a massless 4-point scalar amplitude $\M$ is a function of Mandelstam variables $s,\,t,\,u$.  With the relation $s+t+u=0$, there are only two independent kinematic variables, which can be chosen as the center-of-mass energy $\omega$ and an angular variable $z$, given by\footnote{We use the  convention for $z$ in Ref.~\cite{Arkani-Hamed:2020gyp}, which is different from the one in Ref.~\cite{Chang:2021wvv}. The two conventions are related by $z \leftrightarrow 1/z$.  We always assume $s$ to be the physical channel with $s\geq 0$ and $u,\,t\leq 0$.} 
\begin{equation}
s=\omega^2\,, \hspace{1cm} t = -z\omega^2\,, \hspace{1cm}  u = -(1-z) \omega^2 \,.
\end{equation}
with $0\leq z\leq 1$. The celestial amplitude $\A$ in terms of the boost weight $\beta$ and the angular variable $z$ is then given by the Mellin transform of the amplitude $\M$ with respect to $\omega$,
\begin{equation}
\A(\beta,\, z) = \int^\infty_0  \frac{d\omega}{\omega} \omega^{\beta} \M(\omega, \, z) \,.
\end{equation}
Depending on the form of $\M(\omega, \, z)$, the integral only converges in certain regions of $\beta$.  One could nevertheless perform analytic continuation on $\A(\beta,\, z)$ so that it is defined on the entire complex $\beta$-plane.  It is the analytic structure of $\A(\beta,\, z)$ on the $\beta$-plane that we wish to examine in this paper.  For amplitudes with spinning particles, an additional factor is needed to characterize its little group scaling, while the celestial amplitude is still obtained by the same procedure~\cite{Pasterski:2017ylz, Arkani-Hamed:2020gyp}. 

Several important observations were made in Ref.~\cite{Arkani-Hamed:2020gyp} about the analytic structure of celestial amplitudes.  The analytic structure in the negative (positive) $\beta$-plane generally corresponds to the physics in the deep IR (UV).  In the deep IR, one could expand the amplitude in terms of $\omega$ and obtain an EFT.  
At tree level, it is simply given by a polynomial of $\omega$, which gives simple poles at negative integers in the $\beta$-plane after Mellin transformation. As the residues of these simple poles are given by the corresponding Wilson coefficients, they satisfy various positivity constratins implied from causality and unitarity~\cite{Adams:2006sv,deRham:2017avq, Correia:2020xtr, Tolley:2020gtv, Guerrieri:2020bto, Caron-Huot:2020cmc, Arkani-Hamed:2020blm}.
On the other hand, the analytic structure in the positive $\beta$-plane depends on the nature of the UV theory. Poles at positive integers are expected for a general field theory; while in quantum gravity, the positive $\beta$-plane is completely analytic.  Loop contributions in general massless EFT can be expressed in terms of a series in $\log\omega$, which generates higher order poles in the negative $\beta$-plane. 

Let us be more specific on the statements for EFTs.  A 4-point amplitude in a general massless EFT can be written as 
\begin{equation}
\M(\omega) = \underset{a,b \geq 0}{\sum} c_{a,b} \omega^a \log^b{\omega} \,,  \label{eq:aeft}
\end{equation}
where the coefficient $c_{a,b}$ are functions of $z$ (which is not explicitly written).  The parameters $a$ and $b$ correspond to the EFT expansion and the loop expansion, respectively.  
The EFT is only valid in the low energy region. As such, we will only look at the low energy contribution, cutting off the integration at some arbitrary energy $\omega_0$, assuming it is smaller than the scale of the possible UV physics.  
Without loss of generality, we can set $\omega_0=1$ and obtain\footnote{The integration is obtained in the region $\beta+a >0$  where it converges, and analytically continued to the entire $\beta$-plane. 
We have checked that a more general $\omega_0$ does not generate additional analytic structures.}
\begin{equation}
\A (\beta) = \underset{a,b \geq 0}{\sum} c_{a,b}  \int^1_0  \frac{d\omega}{\omega} \omega^{\beta+a}  \log^b{\omega} 
=\underset{a,b \geq 0}{\sum}   \frac{c_{a,b} (-1)^b \, b!}{(\beta+a)^{b+1}} \,,  \label{eq:mel1}
\end{equation}
which has a pole at $-a$ of order $b+1$ for any given $a$ and $b$. 
Therefore, the order of the EFT expansion (parameterized by the power of $\omega$) determines the position of the pole, which are at negative integer values on the $\beta$-plane, while the order of the loop correction (parameterized by maximum power of the $\log\omega$ term) determines the order of the pole. Notice that the above integration has to include physics in the $\omega\rightarrow 0$ limit to manifest these pole structures. With any finite IR cut-off, taking $\beta\to -a$ gives finite $\mathcal{A}(\beta)$.

Unfortunately, the statements above have a critical flaw: as shown in \autoref{eq:mel1}, for each given $a$, the contribution to the celestial amplitude is given by a series of poles at $\beta = -a$ in the form $\underset{b \geq 0}{\sum} \frac{f(b)}{(\beta+a)^{b+1}}$.  In the region sufficiently close to $\beta = -a$, the terms with higher order poles always dominate, and the perturbative loop expansion breaks down.  
In this case,  the results at a fixed loop order generally would not capture the correct analytic structure.  As a naive example, let us consider the Laurent series of the following expression with a small $\xi$, defined in the region outside the point $-a+\xi$,
\begin{equation}
\label{eq:genshift}
\frac{1}{\beta+a - \xi} = \frac{1}{a+\beta} + \frac{\xi}{(a+\beta)^2} +  \frac{\xi^2}{(a+\beta)^3} + \frac{\xi^3}{(a+\beta)^4} + ... \,, 
\end{equation}
which can be considered as a loop expansion with $\xi$ being the loop suppression factor ($\sim 1/16\pi^2$). The lefthand side has a simple pole at $-a+\xi$, while the righthand side contains a series of higher order poles at $-a$.  The expansion breaks down in the region $|\beta+a| < |\xi|$, and fails to capture the analytic structure in this region!

Indeed, this ``non-perturbativity'' is inherited from the amplitude in \eq{aeft}, by noting that the analytic structure in the negative $\beta$-plane is dominated by the physics in the deep IR region, $\omega\rightarrow 0$.  It is well known that, the $\log \omega$ terms become sufficiently large in this region, and higher order terms in $\log \omega$ are more important and need to be resummed.  With resummation, indeed the large log problem is resolved, and the amplitude stays valid even in the $\omega \to 0$ limit, unless the couplings become sufficiently large at the IR to reach a confinement scale ({\it e.g.} as in QCD).
As a first step, we will consider the resummation of leading log contributions using the method of renormalization group (RG), assuming no confinement in the IR.  
We will show that, at this order, in the massless EFT, the analytic structures of the celestial amplitude are already much richer than the multi-pole structure, and may contain shifted simple poles (as in the example above) or, more generally, branch cuts.  A better understanding of these structures could thus provide important insights on the analytic structures of celestial amplitudes in general, and may eventually lead to a deeper understanding of quantum field theory.  
Of course, it is important to check the effects of sub-leading log contributions, {\it i.e.} whether they can generate new analytic structures.  As we will discuss later, these contributions bring mathematical challenges in the calculation of celestial amplitudes. New approaches are likely needed to make further progress.

The rest of this paper is organized as follows: We begin in \secref{secii} by discussing the general structures of massless EFT amplitudes with resummation of the leading log contributions, focusing on the $\omega$ dependence which is most important for the celestial amplitudes. We then discuss in \secref{seciii} their corresponding celestial amplitudes below a certain threshold. With the simple case of considering one higher dimensional operator renormalized by a renormalizable operator, general properties of celestial amplitude are presented. The discussions are further extended to the most general case with operators at different dimensions. 
In \secref{seciv}, we give an example of a four-scalar scattering amplitude, followed by an interesting case of a four-fermion scattering amplitude that exhibits accidental multipole structures at the leading-log order. 
Our conclusion is drawn in \secref{secv}.

\section{Resummed EFT amplitudes}
\label{sec:secii}

Resummation of leading log contributions can be efficiently done in EFTs via the RG running of operator coefficients (see \cite{manohar2018introduction} for a recent review).  Starting from the UV theory, the standard procedure is to integrate out the heavy resonances and match it to the EFT at some matching scale $M$, which generates a set of operator coefficients.  These coefficients are then RG run down to a lower scale of interest $\mu$ ({\it e.g.} the scale of experimental measurements).  If the scale separation is large, the higher order terms in $\log\frac{\mu}{M}$ become non-negligible and need to be resumed.  As we argued in the previous section, this resummation must be included in the amplitudes in order to obtain the accurate analytic structures of celestial amplitudes. The inclusion of the resummation in the EFT amplitudes follows from a well known procedure, facilitated by the use of RGEs. In the rest of the section, for completeness and establishing our notation, we review this procedure and derive the result in our context for later use.

Given an EFT Lagrangian, it is straight forward to calculate the amplitude of a process with field theory procedures.  However, since our main objective is to Mellin transform an amplitude, it is more illustrative to adopt the on-shell approach and directly parameterize EFTs in terms of amplitudes~\cite{Shadmi:2018xan, Aoude:2019tzn,Ma:2019gtx,2020,Dong:2021yak,AccettulliHuber:2021uoa}.  Recently, it was shown that this approach also provides an efficient way of calculating the anomalous dimension matrices and general pattern of possible loop effects in EFTs~\cite{2015, Bern:2019wie, Craig:2019wmo,Bern:2020ikv, Baratella:2020lzz}.  In a massless EFT, the mapping between higher dimensional operators and the amplitudes is straight forward.  
We will again use a 4-point massless scalar amplitude for illustration.  The generalization to other cases is discussed at the end of this section.  

At tree level, the 4-point amplitude can be written as an expansion of the Mandelstam variables (assuming there is no 3-point renormalizable scalar interaction), 
\begin{equation}
\M(s, t) = c_0 + c^{(0)}_{2} \, s + c^{(1)}_{2} \, t  + c^{(0)}_{4} \, s^2 + c^{(1)}_{4} st + c^{(2)}_{4} \, t^2 + ... \,,  \label{eq:amp0}
\end{equation}
where we have used the relation $s+t+u=0$ to eliminate $u$. $c_0$ is a dimensionless coupling while $c^{(i)}_{2}$ ($c^{(i)}_{4}$) has mass dimension $-2$ ($-4$) and corresponds to dimension-6 (8) operator coefficient. In general, each independent kinematic term has an independent coefficient (with different labels on the superscripts), while symmetries may impose additional relations among them.  Writing $\M$ in terms of $\omega$ and $z$, we have
\begin{equation}
\M(\omega, z) = c_0 + \left( c^{(0)}_{2} - z c^{(1)}_{2}  \right) \omega^2 + \left( c^{(0)}_{4} - z  c^{(1)}_{4} + z^2 c^{(2)}_{4} \right) \omega^4 + ... \,,   \label{eq:amp1}
\end{equation}
where for each order in the $\omega$ expansion, different kinematic terms are parameterized by different powers of $-z$.  Note that we have chosen the superscripts of the $c$ coefficients to match the power of $-z$.

As shown in \autoref{eq:aeft}, loop contributions bring in $\log\omega$ terms in the amplitudes.  
Instead of the log expansion in \autoref{eq:aeft}, we assume the amplitude can be written as
\begin{equation}
\M(\omega, z) = \underset{a \geq 0}{\sum} f_a(c^{(i)}_a, \, \log \frac{\omega}{\mu}, \, z) \, \omega^a \,,  \label{eq:aeft2}
\end{equation}
where $c^{(i)}_a$ are now the running Wilson coefficients that depend on the renormalization scale $\mu$.  
Keeping only the contributions up to one loop, $f_a(c^{(i)}_a, \, \log \frac{\omega}{\mu}, \, z)$ can be written in the form
\begin{equation}
\label{eq:ffun}
f_a 
=\underset{i = 0}{\overset{a/2}{\sum}}  (-z)^i \left(c^{(i)}_a + \gamma_{ij} \bar{c}^{(j)}_{a}  \log \frac{\omega}{\mu} + ...  \right), 
\end{equation}
where 
$\bar{c}^{(j)}_a$ are combinations of couplings that enter the loop (which has the same dimension as $c^{(i)}_a$), 
and the ``...'' part contain additional terms that are independent of $\mu$ at one loop order, which include possible $\log(z)$ and $\log(1-z)$ terms from loop kinematics\footnote{For instance, $\log\frac{-t}{\mu^2} = \log\frac{\omega^2}{\mu^2} + \log(z)$. Note that, in the forward ($z\rightarrow 0$)  or backward ($z\rightarrow 1$) limit, the logarithmic factor $\log(z)$ or $\log(1-z)$ diverges and the amplitude may contain IR divergences.  
Conventionally, we choose arguments of complex numbers to be in the interval $\left(-\pi, \pi\right]$. The corresponding branch of the complex logarithm has discontinuities all along the negative real x axis}.
We have also absorbed factors of $1/16\pi^2$ into $ \gamma_{ij}$.  At this point, \autoref{eq:aeft2} with \autoref{eq:ffun} is simply \autoref{eq:aeft} truncated to the order $b\leq 1$.  However, writing in the form of \autoref{eq:ffun} makes it particularly convenient to derive the RGEs.  Given that $\omega$ and $z$ are kinematic variables (which can vary), for the physical amplitudes to be independent of $\mu$, each $\left(c^{(i)}_a + \gamma_{ij} \bar{c}^{(j)}_{a}  \log \frac{\omega}{\mu} + ...  \right)$ term must be separately independent of $\mu$,  
which gives a one-loop RGE for each $c^{(i)}_a$ as 
\begin{equation}
\frac{d \, c^{(i)}_a}{ d \log\mu} = \gamma_{ij} \bar{c}^{(j)}_a \,.  \label{eq:rgci}
\end{equation}
Note that,  the $\mu$ dependence in $\bar{c}^{(j)}_a$ can be neglected at the one loop order.  These RGEs then captures the $\log\omega$ dependence of the (re-summed) one loop contribution. 
In particular, the solution of \autoref{eq:rgci}, when expanded to one loop order ({\it i.e.} not re-summed), is given by 
\begin{equation}
c^{(i)}_a(\mu) = c^{(i)}_a(\mu_0) + \gamma_{ij}  \bar{c}^{(j)}_a \log \frac{\mu}{\mu_0} \,,
\label{eq:crun}
\end{equation}
where $\mu_0$ is some reference scale, often chosen to be the matching scale $M$ where $c^{(i)}_a(M)$ is calculated from the UV theory. 
Substituting \eq{crun} into \eq{ffun} with $\mu_0=M$, we have
\begin{equation}
\label{eq:ffun2}
f_a 
=\underset{i = 0}{\overset{a/2}{\sum}}  (-z)^i \left(c^{(i)}_a(M) + \gamma_{ij} \bar{c}^{(j)}_{a}  \log \frac{\omega}{M} + ...  \right), 
\end{equation}
which is indeed independent of the renormalization scale $\mu$ at the one-loop order. 
Another way to obtain \autoref{eq:ffun2} is to set the renormalization scale $\mu\to g(z) \omega$, with $g(z)$ being an order one factor.\,\footnote{The factor $g (z)$ parameterizes additional $z$ dependence that may come from the loop kinematics.} This way, the log term $\log \omega/\mu$ in \eq{ffun} can be neglected, and we get $f_a = \underset{i}{\sum} (-z)^i c^{(i)}_a(\omega) + ...$ where $c^{(i)}_a(\omega)$ is given by \autoref{eq:crun} with the substitution $\mu\to g(z) \omega$.  
To effectively resum potential large logs, we will consider $g(z)$ to be order one factor in the following discussions. Importantly, the last method can be generalized to the resummed case, where the amplitude is given by 
\begin{equation}
\M(\omega) = \underset{a, i}{\sum}  (-z)^i c^{(i)}_a(\mu) \omega^a  |_{\mu\to g(z) \omega} \,,   \label{eq:aeft3}
\end{equation}
and $c^{(i)}_a(\mu)$ is given by the exact solution of the RGE \autoref{eq:rgci} instead of the one-loop result \autoref{eq:crun}. 
Note that \autoref{eq:aeft3} omits the rational loop contributions which are expected to be sub-leading.  For a general massless EFT, the rational loop terms will depend on $\omega$ polynomially and will contribute to single pole structures of the celestial amplitude. When including non-vanishing mass for the light particles, amplitudes at one loop level also include branch-cut for producing massive-particles at threshold. The analytical structure of the corresponding celestial amplitude is an interesting topic to be explored for the future.
\autoref{eq:aeft3} is our master formula for calculating the RG-resumed amplitude.  

It is straight forward to apply the above procedure to spinning particles.  The amplitude, written in terms of $\omega$ and $z$, contains an additional factor of spinor products from the little group scaling.  The detailed derivation of this factor can be found in \autoref{sec:celesphere} and Ref.~\cite{Arkani-Hamed:2020gyp}.  A 4-point amplitude may also contain massless poles generated by on-shell 3-point amplitudes, though the existence of such 3-point amplitudes are subject to the operator dimensions and little group scalings.\footnote{For instance, for particles with spin $\leq 1$, the only on-shell 3-point amplitude at the level of dimension-6 operators is the 3-vector one with same helicities.  See {\it e.g.} Refs.~\cite{2015, Craig:2019wmo} for more details.}  In general, one could replace the $(-z)^i$ factors in \autoref{eq:aeft3} with some more general rational functions of $z$.  As long as one chooses a ``non-redundant basis'' for such $z$ functions to parameterize the $f_a$ in \autoref{eq:ffun}, the above derivation still holds and a RGE can be written down for each $c^{(i)}_a$.


\section{Celestial Amplitude}
\label{sec:seciii}

With the prescription in \autoref{eq:aeft3} we are now ready to calculate the leading-log resumed amplitude of a given theory and obtain its celestial amplitude.  Let us start with the simplest nontrivial case in \autoref{sec:celei}, where an amplitude receives contributions from both a dimensionless coupling and an irrelevant coupling (Wilson coefficient).  This case already contains all the essential features of the analytic structures.  We then move on to a more complicated case in \autoref{sec:celeii} where a series of Wilson coefficients are considered.  

\subsection{Wilson Coefficient Running from Dimensionless couplings}
\label{sec:celei}

We consider the following 4-point amplitude with two running couplings $\lambda(\mu)$ and $c_a(\mu)$
\begin{equation}
\M(\omega) = \lambda(\mu) + c_a(\mu) \omega^a + ... \bigg|_{\mu\to g(z)\omega}\,,
\end{equation}
where for simplicity, we have omitted the possible power-law $z$ dependences, which is not relevant for the following discussion. $\lambda$ is a dimensionless coupling while $c_a$ has mass dimension $-a$.  A typical example of this is the amplitude of a complex scalar $\phi\phi \to \phi\phi$, where the only dimension-6 operator contribution is given by $c_2 \omega^2$ due to the $t\leftrightarrow u$ symmetry of the amplitude.  Here, we will work in a general framework without referring to any particular model.

The one-loop RGEs of $\lambda(\mu)$ and $c_a(\mu)$ are given by
\begin{align}
\frac{d\, \lambda}{d\log\mu}  =&~ \glam \lambda^2 \,, \label{eq:rgelam} \\
\frac{d\, c_{a}}{d\log\mu} =&~ \gc  c_{a} \lambda \,,   \label{eq:rgeca}
\end{align}
where $\glam$ and $\gc$ are determined by the particular theory and contain a loop factor ($\sim 1/16\pi^2$).  Note that the form of these RGEs is dedicated by dimensional analysis. In a massless theory, $\lambda(\mu)$ does not receive contributions from $c_a$.  One could solve \autoref{eq:rgelam} first, which gives
\begin{equation}
\lambda(\mu) = \frac{\lambda_M}{1-\glam\lambda_M\log\frac{\mu}{M}}  \,,  \label{eq:sollam}
\end{equation}
where we have conveniently fixed the boundary condition at the matching scale, $\lambda_M \equiv \lambda(\mu=M)$.
We will focus on the case with $\glam \geq 0$ so that the dimensionless coupling $\lambda(\mu)$ is finite in the IR region $\omega\rightarrow 0$ ({\it i.e.} no confinement).  On the other hand, $\gc$ can be either positive or negative, as the $\omega^a$ term always dominates over any log divergences to make the EFT contribution well-behaved ({\it i.e.} irrelevant) in the IR.

Let us first consider a special case in which $\glam \rightarrow 0$, {\it i.e.} $\lambda$ does not run.  The solution to \autoref{eq:rgeca} in this case is given by
\begin{equation}
c_a (\mu) = c_{a_M} \left(\frac{\mu}{M}\right)^{\gc\lambda_M}  \,,  \label{eq:solca1}
\end{equation}
where we have again fixed the boundary condition at the matching scale $M$ (with $c_{a_M}\equiv c_a(\mu=M) $).   Following the prescription in \autoref{eq:aeft3}, the RG-improved amplitude is given by 
\begin{equation}
\M(\omega) = \lambda_M + c_{a_M} \left(\frac{\omega}{M}\right)^{\gc\lambda_M} \omega^a  + ... \,, \label{eq:ampomg1}
\end{equation}
where for simplicity we have set $g(z)=1$, since it generates an overall factor which does not change the analytic structure   of celestial amplitudes.  
The $\lambda_M$ contribution generates a simple pole at $\beta=0$ in the celestial amplitude as in the tree-level case~\cite{Arkani-Hamed:2020gyp}. 
For the contribution of $c_a$, we have
\begin{align}
\label{eq:mellinshift}
\mathcal{A} (\beta) =&~ \int^M_{0} c_{a_M} \bigg(\frac{\omega}{M}\bigg)^{\gc\lambda_M} \omega^a\omega^{\beta-1} d\omega  \nonumber\\
=&~ c_{a_M} \frac{M^{a+\beta}}{\beta + a+\gc\lambda_M} \,,
\end{align}
where we have naturally chosen to cut off the integration at $M$ as well.\footnote{Again, a different cut off generates an overall factor which does not affect the analytic structure.}  
The integration is obtained in the region $\beta+a+ \gc\lambda_M > 0$ where it converges, and analytically continued to the entire $\beta$-plane.  
We see that, the celestial amplitude has a simple pole at $\beta = -a-\gc \lambda_M$, which is shifted from the tree-level simple pole at $\beta =-a$.  This is exactly the case described by \autoref{eq:genshift}.  Without resummation, the series in $\log{\omega}$ would generate a series of multipoles at $\beta =-a$, which does not capture the correct analytic structure.  Note that the shifted-pole observed here for an IR finite amplitude is different from the one related to IR-divergent amplitudes as found in \cite{Arkani-Hamed:2020gyp}. 

The case of nonzero $\glam$ is even more interesting.  In this case, the solution to \autoref{eq:rgeca} is
\begin{equation}
c_a (\mu) = c_{a_M} \left( 1 -  \glam \lambda_M \log\frac{\mu}{M}  \right)^{-\clam}  \,. \label{eq:solca2}
\end{equation}
Note that, \autoref{eq:solca2} reduces to \autoref{eq:solca1} in the limit $\glam \to 0$ as expected.  The RG-improved amplitude is given by 
\begin{align}
\M(\omega) =&~ \lambda_M \left(1-\glam\lambda_M\log\frac{g(z)\omega}{M}\right)^{-1} + c_{a_M} \left( 1 -  \glam \lambda_M \log\frac{g(z)\omega}{M}  \right)^{-\clam} \omega^a  +  ...  \nonumber\\ 
=&~ \lambda_M \kappa^{-1} \left(\rho - \log\frac{\omega}{M} \right)^{-1} + c_{a_M} \kappa^{-r} \left(\rho - \log\frac{\omega}{M} \right)^{-r}  \omega^a  +  ... \,,  \label{eq:ampomg2}
\end{align}
where in the second line we have defined $r = \clam$, $\kappa = \gamma_\lambda \lambda_M$ and $\rho = \kappa^{-1} - \log g(z)$ to simplify the expressions. As $g(z)$ being an order-one factor, we will restrict ourselves to the case of $\rho > 0$.

Focusing on the $c_a$ contribution, the Mellin transformation of the amplitude is given by:
\begin{align}
\A(\beta) =&~  c_{a_M}   \kappa^{-r} \int^M_{0} \left(\rho - \log\frac{\omega}{M} \right)^{-r} \omega^{a+ \beta-1} d\omega \nonumber\\ 
=&~   c_{a_M}   \kappa^{-r} M^{\beta+a} e^{\rho(\beta+a)} (\beta+a)^{r-1}  \int_{\rho(\beta+a)}^{\infty} e^{-t'} t'^{-r} dt' \,, \nonumber \\
=&~  c_{a_M} \kappa^{-r} M^{\beta+a} e^{\rho(\beta+a)} (\beta+a)^{r-1} \Gamma(1-r, \rho (\beta+a)) \,,   
\label{eq:mellinoneloop}
\end{align}
where in the second line we have performed changes of variables $t =\log \frac{\omega}{M}$ and then $t' = (\beta+\alpha)(\rho-t)$, and in the last line we used the definition of the incomplete Gamma function, 
$\Gamma(s,x) = \int_x^\infty t^{s-1} e^{-t}dt$.  Note again that the integration is obtained in the $\beta > -a$ region
and analytically continued to the entire $\beta$-plane. 
Properties of the incomplete Gamma function~\cite{wikiInGamma} lead to a few important observations. 

\begin{figure}[h!]
\includegraphics[scale=0.15]{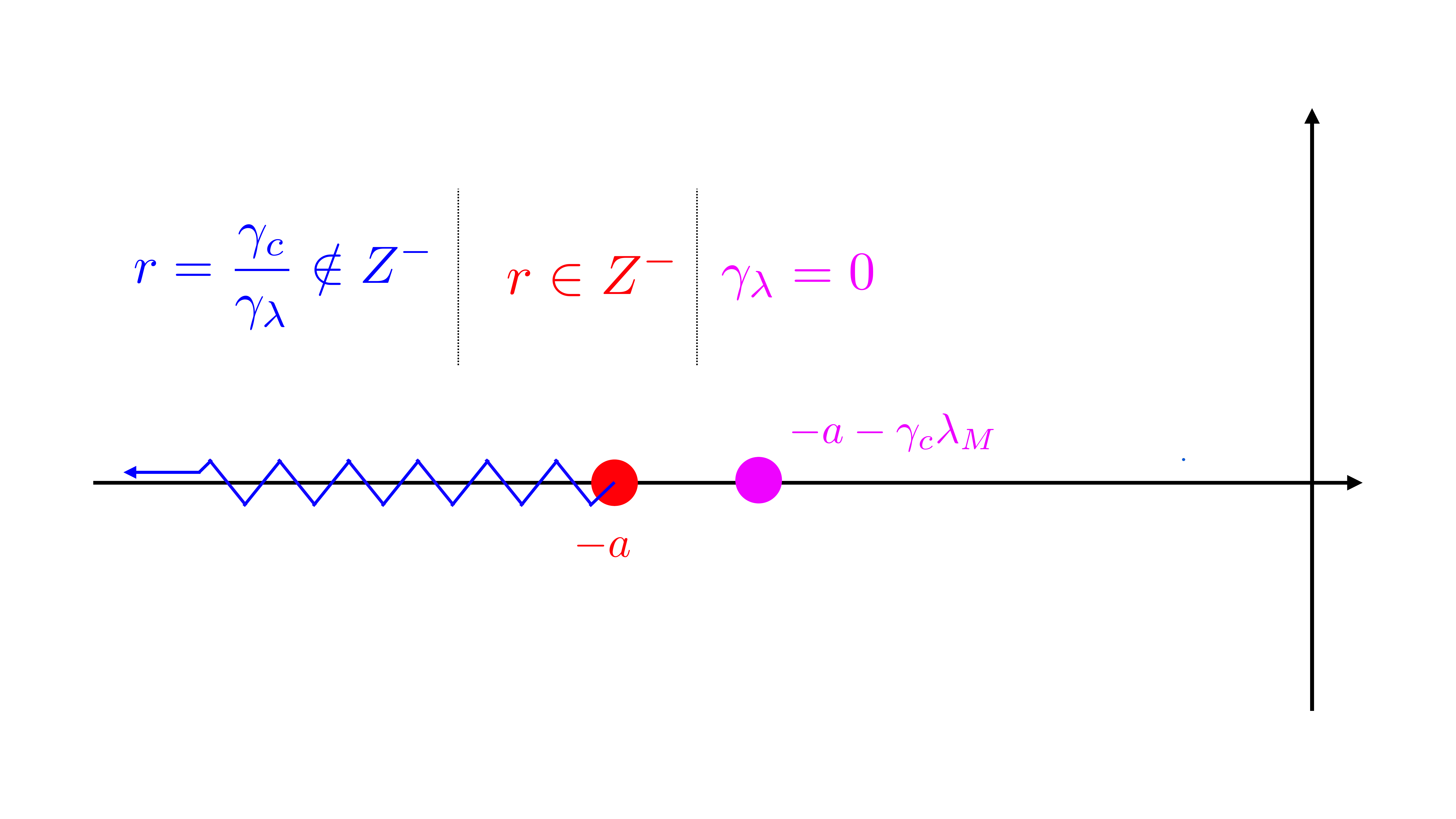}
\caption{Analytical structures of the celestial amplitudes considering one-loop resummation. The shifted pole position depends on the sign of $\gamma_c$, where for clarity we show the case with negative $\gamma_c$.}
\label{fig:celes}
\end{figure}

  In general, $\A(\beta)$ has a branch cut on $(-\infty, a]$.  Furthermore, we find that at $\beta = -a$, $\A(\beta)$ is finite (divergent) for $r > 1$ ($r<1$), and logarithmically divergent for $r=1$.  This dependence on $r$ is related to how fast $c_a$ goes to zero or diverges as $\omega\to 0$, as shown in \autoref{eq:solca2}. More details of these results can be found in \autoref{sec:Gamma}.

We also note that, for small non-zero $\glam$, (the first line of) \autoref{eq:ampomg2} can be expanded in $\glam$ which reproduces \autoref{eq:ampomg1} with a series of higher orders terms in $\log\omega$, and the celestial amplitude has a series of multipoles at $\beta =-a-\gc \lambda_M$.  Expanding \autoref{eq:ampomg2} in terms of $\log\frac{\omega}{M}$ instead, one then recovers the series in \autoref{eq:aeft} with the celestial amplitude having multipoles at $\beta =-a$.  In both cases, the expanded celestial amplitude could not reproduce the analytic structure of the original one.  As explained earlier, this is because the $\log\frac{\omega}{M}$ series diverges in the deep IR $\omega\to0$, which the analytic structures of celestial amplitudes are sensitive to. 

We would also like to point out an interesting special case.    Note that $\Gamma(s,x) = (s-1)!e^{-x} \underset{k=0}{\overset{s-1}{\sum}} \frac{x^k}{k!}$ if $s$ is a positive integer. Therefore, when $r$ is a nonpositive integer, we have 
\begin{equation}
\A (\beta) = c_{a_M} \kappa^{-r} M^{\beta+a} (-r)!  \underset{k=0}{\overset{-r}{\sum}} \frac{\rho^k}{k! (\beta+a)^{1-r-k}} \,,
\end{equation}
which contains a series of multipoles up to order $1-r$. This is also expected, since in this case the $c_a$ contribution in \autoref{eq:ampomg2} can be expanded, and the log series actually terminates at the order $-r$. For $r=0$, we thus have $\A(\beta) = \frac{ c_{a_M} M^{\beta+a} }{\beta +a }$.  This is exactly the tree-level result since $c_{a_M}$ does not run when $r=0$. A negative integer $r$ is generally difficult to obtain, but we find it possible to tune the particle contents and the corresponding quantum numbers for certain theories in a way that negative integer $r$ can be realized at the leading-log level.  In this case, the higher order log contributions accidentally cancel.  In \secref{seciv} we will show an explicit example of it.

    To summarize, we observe the following structures for $\A(\beta)$ on the complex $\beta$-plane as shown in Fig.~\ref{fig:celes}:
\begin{itemize}
\item $r \in Z^-$: poles at $-a$ with pole order up to $1-r$. 
\item $\gamma_\lambda = 0~(r\to \infty)$: shifted poles at $-a -\gamma_c\lambda_M$.
\item for other values of $r$: branch cut on $(-\infty, a]$.
\end{itemize}


\subsection{Running of Higher Dimensional Operators}
\label{sec:celeii}

Let us now apply the analysis in \autoref{sec:celei} to a more general amplitude with a series of Wilson coefficients as in \autoref{sec:secii}.  For simplicity, we only consider 4-point interactions, and denote $c_0$ as $\lambda$, a common notation for a 4-point dimensionless scalar interaction.  Note however that our RGE results below also apply if the renormalizable part of the amplitude is constructed by two 3-point interactions ({\it e.g.} for 4-fermion amplitudes) with the replacement $\lambda \to g^2$ (gauge coupling) or $\lambda \to y^2$ (Yukawa coupling). Indeed, the RGEs of $\lambda$, $g^2$ and $y^2$ in the SM all have the same structure. On the other hand, if the anomalous dimension matrices involve different dimensionless couplings, e.g. operator mixings between $\psi^2 X H$ and $X^2 H^2$ where the dimensionless four-point coupling part is proportional to $gy$~\cite{Jenkins:2013zja}, the structures of RGEs below could become more complicated and difficult to solve.  We do not consider these cases here. 

With the above assumptions, the general one-loop RGE of the Wilson coefficient $c^{(i)}_a$ can be written as (the indices $j,k$ are summed over)
\begin{equation}
\frac{d c^{(i)}_a}{ d \log \mu} = \gamma_{a}^{ij} c^{(j)}_a \lambda + \underset{a_1\geq 1, a_2\geq 1}{\overset{a_1+a_2 =a}{\sum}} \gamma_{a,a_1,a_2}^{ijk} c^{(j)}_{a_1} c^{(k)}_{a_2}  \,, \label{eq:rgci2}
\end{equation}
where, instead of \autoref{eq:rgci}, the righthand side of the RGE is written in two parts, the first part characterize  the RG mixing among the $c^{(i)}_a$s, and the second part are the contributions from lower-dimensional coefficients.  We can always choose the basis of $c^{(i)}_a$ to diagonalize the matrix $\gamma_{a}^{ij}$, so that \autoref{eq:rgci2} can be written as 
\begin{equation}
\frac{d c^{(i)}_a}{ d \log \mu} = \gamma_{a}^{(i)} c^{(i)}_a \lambda + \underset{a_1\geq 1, a_2\geq 1}{\overset{a_1+a_2 =a}{\sum}} \gamma_{a,a_1,a_2}^{ijk} c^{(j)}_{a_1} c^{(k)}_{a_2}  \,. \label{eq:rgci3}
\end{equation}
\autoref{eq:rgci3} can be solved for all $c^{(i)}_a$ by induction.  First, we note that for $\lambda$ and $c^{(i)}_1$,\footnote{Or the lowest-order non-zero $c^{(i)}_a$, for instance $c^{(i)}_2$ in the scalar case in \autoref{eq:amp1}.  
} the second term on the righthand side is absent and the solutions are given by (see \autoref{eq:sollam} and \autoref{eq:solca2})
\begin{align}
\lambda(\mu) =&~ \lambda_M \left(1-\glam\lambda_M\log\frac{\mu}{M} \right)^{-1}   \,,  \nonumber\\
c^{(i)}_1 (\mu) =&~ c^{(i)}_{1_M} \left( 1 -  \glam \lambda_M \log\frac{\mu}{M}  \right)^{-\gamma^{(i)}_{1} / \glam}  \,.
\end{align}
Then, the RGE of $c^{(i)}_2$ is given by
\begin{equation}
\frac{d c^{(i)}_2}{ d t} = \gamma_{2}^{(i)} c^{(i)}_2  \lambda_M \left(1-\glam\lambda_M \,t \right)^{-1} +  \gamma_{2,1,1}^{ijk} c^{(j)}_{1_M} c^{(k)}_{1_M}  \left( 1 -  \glam \lambda_M \,t  \right)^\frac{-\gamma^{(j)}_{1}-\gamma^{(k)}_{1}}{\glam}   \,, \label{eq:rgci4}
\end{equation}
where we have defined $t\equiv \log \mu/M $ for convenience.  The solution is 
\begin{equation}
c^{(i)}_2 (t) = A \left(1-\glam\lambda_M \,t \right)^{-\frac{\gamma^{(i)}_{2}}{\glam}} - \frac{ \gamma_{2,1,1}^{ijk} c^{(j)}_{1_M} c^{(k)}_{1_M} \left(1-\glam\lambda_M \,t \right)^{1-\frac{\gamma^{(j)}_{1} + \gamma^{(k)}_{1}}{\glam}}  }{\lambda_M( \gamma^{(i)}_{2} + \gamma_\lambda  - \gamma^{(j)}_{1} - \gamma^{(k)}_{1}   )}   \,,  \label{eq:solc2}
\end{equation}
where the constant $A$ can be fixed by the boundary condition $c^{(i)}_2 (t=0) = c^{(i)}_{2_M} $, and is not explicitly written.  From \autoref{eq:solc2} we could then derive the running of $c^{(i)}_3$, and so on.  Indeed, the RGEs of $c^{(i)}_a$ have the general form 
\begin{equation}
\frac{d c}{d t} = \alpha c \left( 1- \kappa t  \right)^{-1} + \underset{n}{\sum} \beta_n \left( 1- \kappa t  \right)^{-x_n}  \label{eq:crge},
\end{equation}
with $x_n$ being functions of several different $\gamma_{a}^{(j)}/\gamma_\lambda$'s, and $\kappa =\gamma_\lambda \lambda_M$, $\alpha = \gamma_c \lambda_M$. Assuming $x_n -1 -r \neq 0$ for $r= \frac{\gamma_c}{\glam} = \frac{\alpha}{\kappa}$, it has the solution\footnote{When $r+1-x_n =0$, the RGE is solved by replacing the factor $1/(r+1-x_n)$ by $\log(1-\kappa t)$ in \autoref{eq:csol}. We will not consider this case since the corresponding celestial amplitude is difficult to obtain analytically.}
\begin{equation}
c(t) = A \left( 1- \kappa t  \right)^{-r} - \underset{n}{\sum}  \frac{ \beta_n \left( 1- \kappa t  \right)^{1 -x_n} }{ \kappa (r+1-x_n) } \,.  \label{eq:csol}
\end{equation}
With \autoref{eq:csol} and our prescription in \autoref{eq:aeft3}, the amplitude is given by a series of the form $\sum f_{a,i}  \left(\rho - \log\frac{\omega}{M} \right)^{-r_i} \omega^{a+ \beta-1}$, where the Mellin transform of each term is given by \autoref{eq:mellinoneloop} with the replacement $r \to r_i$.  The analytic structures discussed in \autoref{sec:celei} thus still apply in the more general case.  
Let us verify this with an explicit example where the amplitude is given by
\begin{equation}
\M(\omega) = \lambda + c_2 \omega^2 + c_4 \omega^4 + c_{6} \omega^6 + ... \,,
\end{equation}
where, for simplicity, we have only considered operator coefficients with even dimensions, and only one  coefficient at each dimension.  The RGEs can be written as (again with $t \equiv \log \mu/M$)
\begin{align}
\frac{d\lambda}{dt} =&~ \glam \lambda^2 \,,  \nonumber \\
\frac{d c_2}{dt} =&~ \gamma_2 \lambda c_2  \,,  \nonumber \\
\frac{d c_4}{dt} =&~ \gamma_4 \lambda c_4 + \gamma'_4 c^2_2  \,, \nonumber \\
\frac{d c_6}{dt} =&~ \gamma_6 \lambda c_6  + \gamma'_6 c_2 c_4  \,,  \nonumber \\
&... \,,  \label{eq:rge246}
\end{align}
and the celestial amplitude is given by 
\begin{align}
\A(\beta) =&~~ 
\frac{e^{\frac{\beta}{\glam}}}{\glam} \Gamma\!\left(0,\frac{\beta}{\glam}\right) 
+ \frac{e^{\frac{\beta+2}{\glam}}}{\glam}  \left(   \frac{\beta+2}{\glam} \right)^{r_2-1} \!\! \Gamma\! \left(1-r_2, \frac{\beta+2}{\glam} \right)  \nonumber \\
&+  \frac{e^{\frac{\beta+4}{\glam}}}{\glam}  \left[ A_4  \left(   \frac{\beta+4}{\glam} \right)^{r_4-1} \!\! \Gamma\! \left(1-r_4, \frac{\beta+4}{\glam} \right) 
+ \frac{ \gamma'_4  \left(   \frac{\beta+4}{\glam} \right)^{2 r_2-2}  \Gamma \!\left(2-2 r_2, \frac{\beta+4}{\glam} \right)}{\glam(2r_2-1- r_4)}  \right] \nonumber \\
&+ \frac{e^{\frac{\beta+6}{\glam}}}{\glam}   \left[  
A_6  \left(   \frac{\beta+6}{\glam} \right)^{r_6-1} \!\!  \Gamma\! \left(1-r_6, \frac{\beta+6}{\glam} \right)  +   \frac{ A_4 \gamma'_6  \left(   \frac{\beta+6}{\glam} \right)^{r_2+r_4-2}  \Gamma \!\left(2-r_2 - r_4 , \frac{\beta+6}{\glam} \right)}{\glam(r_2 + r_4-1 - r_6)} \right. \nonumber\\
& \hspace{1.5cm}+\left.   \frac{  \gamma'_6  \gamma'_4 \left(   \frac{\beta+6}{\glam} \right)^{3 r_2-3}  \Gamma \!\left(3-3 r_2, \frac{\beta+6}{\glam} \right)}{\glam^2(2r_2-1-r_4)(3r_2-2-r_6)}   \right] + ...
\label{eq:celegeneral}
\end{align}
where $r_a = \gamma_a/\glam$, $\lambda(t=0)=1$, $c_2(t=0)=1$, and $A_4$, $A_6$ are some constants fixed by the boundary conditions of $c_4$ and $c_6$.  For simplicity, we have also fixed the $g(z)$ in \autoref{eq:aeft3} to be one.  Several interesting observations can be made for the $\A(\beta)$ in \autoref{eq:celegeneral}.  First, it can be shown that, in the limit $\glam \to 0$, $\A(\beta)$ still reduces to a series of shifted simple poles, following the derivation in the previous section. The poles are shifted by factors depending on only $\gamma_{a}$'s ($\gamma_\lambda,\,\gamma_2,\,\gamma_4,\,...$) which are the anomalous dimension matrices involving $\lambda$, not depending on the ones from only dimensional Wilson coefficients ($\gamma'_4,\,\gamma'_6,\,...$).  Second, for nonzero $\glam$, the analytic structure of $\A(\beta)$ also only depends on $\gamma_{a}$'s, as a result of the structure of the running couplings in \autoref{eq:solc2}. The analytical structure of $\A(\beta)$ is thus sensitive to specific contributions to the RGEs. This pattern should be general if we include only one dimensionless couplings, e.g. $\lambda$, $y$ or $g$. When including multiple dimensionless couplings, the analytical structure of $\A(\beta)$ is difficult to obtain and may have entangled dependence on various anomalous dimensions. Generally, \autoref{eq:celegeneral} contains branch cuts on the negative $\beta$ axis starting from $\beta = 0,\,-2\,,-4\,...$, unless the corresponding parameters ($r_2$, $r_4$, $2r_2-1$, $r_6$, ...) are non-positive integers, as discussed in the previous section.

\section{A Couple of Examples}
\label{sec:seciv}
\subsection{$\lambda \phi^4$}
For simplicity, we will focus on the Higgs sector with the dimensionless coupling $\lambda |H^\dagger H|^2$ in the Standard Model EFT (SMEFT) and consider the four scalar scattering amplitude $\mathcal{A}(H^2_{\alpha\beta}H^{\dagger 2}_{\dot{\alpha}\dot{\beta}})$ with $\alpha, \beta, \dot{\alpha}, \dot{\beta}$ being the indices of $SU(2)_L$. This scattering amplitude receives contributions from the following two dimension-6 operators in the unbroken phase:
\begin{equation}
\mathcal{O}_{H\Box} = (H^\dagger H)\Box (H^\dagger H),~~~~\mathcal{O}_{HD} = (H^\dagger D_{\mu}H)^{*}(H^\dagger D_{\mu}H),
\end{equation}
with the corresponding Wilson coefficients $C_{H\Box}$ and $C_{HD}$, respectively. The anomalous dimension matrix for SMEFT has been computed in~\cite{Jenkins:2013zja, Chetyrkin:2012rz, Jenkins:2013wua, Alonso:2013hga, MACHACEK198383, Roy:2019jqs} with the following RGEs:
\begin{eqnarray}
\mu \frac{d}{d\mu}\lambda &=& 24\frac{\lambda^2}{16\pi^2},\\
\mu \frac{d}{d\mu}C_{H\Box} &=&24 \frac{\lambda(\mu)}{16\pi^2} C_{H\Box}(\mu), \\
\mu \frac{d}{d\mu}C_{HD} &=&12 \frac{\lambda(\mu)}{16\pi^2} C_{HD}(\mu).
\label{eq:renphi4}
\end{eqnarray}
where we also assume massless particles so that the $\lambda$ is only renormalzied by itself and is solved as:
\begin{eqnarray}
\lambda(\mu) = \frac{\lambda_M}{1-24\frac{\lambda(M)}{16\pi^2} \log\frac{\mu}{M}}
\end{eqnarray}
with $\lambda_M = \lambda(\mu = M)$. Solving \eq{renphi4}, we obtain 
\begin{eqnarray}
\label{eq:coefficrunquartic}
C_{H\Box}(\mu)&=& C_{H\Box}(M) \bigg(1 -24\frac{\lambda(M)}{16\pi^2}\log\frac{\mu}{M}\bigg)^{-1} =  C_{H\Box}(M) \bigg(\frac{\lambda(M)}{\lambda(\mu)}\bigg)^{-1}, \\
C_{HD}(\mu) &=& C_{HD}(M) \bigg(1 -24\frac{\lambda(M)}{16\pi^2}\log\frac{\mu}{M}\bigg)^{-1/2} =  C_{HD}(M) \bigg(\frac{\lambda(M)}{\lambda(\mu)}\bigg)^{-1/2}.
\end{eqnarray}
Their contribution to $\mathcal{A}(H^2_{\alpha\beta}H^{\dagger 2}_{\dot{\alpha}\dot{\beta}})$ are given by:
\begin{eqnarray}
\mathcal{A}(H^2_{\alpha\beta}H^{\dagger 2}_{\dot{\alpha}\dot{\beta}}) &\propto & \frac{\omega^2}{M^2}(\delta_{\alpha\dot{\alpha}} \delta_{\beta\dot{\beta}}+ \delta_{\beta\dot{\alpha}} \delta_{\alpha\dot{\beta}})\bigg(C_{HD}(\omega) - 2C_{H\Box}(\omega)\bigg) \notag\\
&+&\frac{(-2z+1)\omega^2}{M^2}(\delta_{\alpha\dot{\alpha}} \delta_{\beta\dot{\beta}} -  \delta_{\beta\dot{\alpha}} \delta_{\alpha\dot{\beta}})\bigg(C_{HD}(\omega) +2 C_{H\Box}(\omega)\bigg),
\end{eqnarray}
where the renormalization scale is set to the collision energy of particles $H^2_{\alpha\beta}$ as $\mu =\omega$. As $\mathcal{A}(H^2_{\alpha\beta}H^{\dagger 2}_{\dot{\alpha}\dot{\beta}})$ receives contributions from the dimension six operator with $r = 1 (C_{H\Box})$ or $r = 1/2 (C_{HD})$, its celestial amplitude diverges at $\beta=-2$ and has branch cut on $\beta \in (-\infty, -2]$.

\subsection{Accidental multipole structures} 
In this subsection, we consider an explicit example where the celestial amplitude generated by an EFT exhibits accidental multipole structures, as mentioned in the previous section.  
For the sake of convenience, we restrict ourselves to dimension-6 operators in the SMEFT, but with strong coupling and the weak couplings turned off in the following discussions.  
In particular, we consider  
the four-fermion scattering amplitudes $\mathcal{A}(L_\alpha Q_{a\beta} u_{\dot{a}}e)$, 
which receives contributions from the two operators
\begin{equation}
\mathcal{O}^{(1)}_{lequ} = (\bar{l}^j e)\epsilon_{jk}(\bar{q}^k u)  \,, \hspace{1.5cm}    \mathcal{O}^{(3)}_{lequ} =  (\bar{l}^j \sigma_{\mu\nu} e)\epsilon_{jk}(\bar{q}^k \sigma^{\mu\nu} u)  \,,
\end{equation}
with the corresponding Wilson coefficients $C^{(1)}_{lequ}$ and $C^{(3)}_{lequ}$, respectively \cite{Grzadkowski:2010es}.  We will fix the hypercharges that fermions carry to be the SM values, and modify them later on to achieve the desired accidental multipole structure. The running of the Wilson coefficients are then given by~\cite{
Alonso:2013hga, MACHACEK198383, Roy:2019jqs}
\begin{eqnarray}
\label{eq:betafun}
\mu \frac{d}{d\mu}C^{(1)}_{lequ} &=&\frac{g^2_1}{16\pi^2} \bigg\{-6 (y^2_e + y_e y_u - y_e y_q + y_q y_u)C^{(1)}_{lequ} - 24 (y_q + y_u) (2 y_e -y_q + y_u) C^{(3)}_{lequ} \bigg\},\notag\\
\mu \frac{d}{d\mu}C^{(3)}_{lequ} &=&\frac{g^2_1}{16\pi^2}\bigg\{-\frac{1}{2}(y_q + y_u) (2y_e -y_q +y_u)C^{(1)}_{lequ}\notag\\
&~&~~~~~~~~~+ 2 ( y^2_e -y_e y_q +y_e y_u - 2y^2_q + 5 y_q y_u - 2y^2_u) C^{(3)}_{lequ}\bigg\},\notag \\
\mu \frac{d}{d \mu}g^2_1 &=&  \frac{n_G g^4_1}{16\pi^2} \frac{4}{3}(2y^2_l + y^2_e + n_c y^2_u + n_c y^2_d + 2n_c y^2_q)\equiv \gamma_g g^4_1,
\end{eqnarray}
where in the SM, $y_q = \frac{1}{6}, y_e = -1, y_d = -\frac{1}{3}, y_u = \frac{2}{3}, y_l = -\frac{1}{2}, n_c = 3$ which gives $\gamma_g = \frac{1}{16\pi^2}\frac{40 n_G}{9}$ (with $n_G$ the number of generations). To solve the differential equations, we can rotate to a basis in which the running of the dimension-six operators is diagonal and can be expressed as:
\begin{eqnarray}
 \mu\frac{d}{d\mu}
 \begin{pmatrix}
 C^1 \\ C^3 
 \end{pmatrix}
 =&& g^2_1\begin{pmatrix}
 \gamma_{c1} & 0 \\
 0 & \gamma_{c3}
 \end{pmatrix}
  \begin{pmatrix}
 C^1 \\ C^3 
 \end{pmatrix}, \notag\\
 ~{\rm with}  &&
 \begin{pmatrix}
 C^{(1)}_{lequ}\\ C^{(3)}_{lequ} 
 \end{pmatrix} = \begin{pmatrix}-\frac{4}{9}(7+2\sqrt{73}) & -\frac{4}{9}(7-2\sqrt{73}) \\ 1& 1\end{pmatrix}  \begin{pmatrix}
 C^1 \\ C^3 
 \end{pmatrix}\equiv  P\begin{pmatrix}
 C^1 \\ C^3 
 \end{pmatrix},
\end{eqnarray}
where $P$ is the invertible matrix to diagonalize the anomalous dimension matrix. The (ratios of) anomalous dimensions turn out to be $\gamma_{c1} /\gamma_g =-0.485...$ and $\gamma_{c3} /\gamma_g =0.226...$, which are irrational numbers. The RGEs can be easily solved, and the solutions follow the pattern in \eq{solca2}:
\begin{eqnarray}
C^1(\mu) &=& C^1(M) \bigg(1-\gamma_g g^2_{1_M} \log\frac{\mu}{M}\bigg)^{-\gamma_{c1}/\gamma_g}, \notag\\
C^3(\mu) &=& C^3(M) \bigg(1-\gamma_g g^2_{1_M} \log\frac{\mu}{M}\bigg)^{-\gamma_{c3}/\gamma_g}.
\end{eqnarray} 
with $g^2_{1_M}\equiv g^2_{1}(\mu=M)$. The four fermion amplitude focusing on the positive-helicity configurations can then be calculated as:
\begin{eqnarray}
&&\mathcal{A}(L^+_\alpha Q^+_{a\beta} u^+_{\dot{a}}e^+) \notag\\
&&~~\propto \epsilon_{\alpha\beta}\delta_{a\dot{a}} \frac{1}{M^2} \bigg(-8 C^{(3)}_{lequ}(\mu) [13][24] -4 C^{(3)}_{lequ}(\mu) [14][23] + C^{(1)}_{lequ}(\mu) [14][23] \bigg) \notag\\
&&~~\propto\frac{\omega^2}{M^2}\bigg\{C^1(\omega)\bigg(P_{11}(1-z)+ 4P_{12}(3z-1)\bigg) + C^3(\omega)\bigg(P_{12} (1-z) + 4P_{22}(3z-1)\bigg)\bigg\},\notag\\
\end{eqnarray}
with the renormalization scale set to $\mu =\omega$. The corresponding celestial amplitude can then be seen to follow our previous discussions, 
and its analytic structure depends on the two important ratios, $\gamma_{c1}/\gamma_g$ and $\gamma_{c3}/\gamma_g$.  
Both ratios will lead to branch-cut structures in the complex $\beta$ plane running from $-\infty$ to $-2$. The  celestial amplitude $\mathcal{A}(\beta)$ is also divergent (both real and imaginary parts) at $\beta = -2$. As $\gamma_{c1} < 0$, $C^1(\omega)$ is divergent in the small $\omega$ region and dominates the contributions to the analytical structure of the celestial amplitudes.

Let us now change the quantum numbers of the fermions in order to obtain the accidental multipole structures.  Following the discussion in the previous section, it is desirable to tune the quantum numbers such that $\gamma_{c1}/\gamma_g$ and $\gamma_{c3}/\gamma_g$ are either zero or negative integers.  
A consistent quantum field theory coupled to gravity may only have fermions with rational hypercharges~\cite{Seiberg, Shiu:2013wxa, Brummer:2009cs}, 
and we will stick to this case.  
Note that, the hypercharges also needs to satisfy the relation $y_l + y_q - y_e - y_u = 0$ to have hypercharge conservation for the two operators, which are already implemented in \autoref{eq:betafun}. 
Consider first the case of $\gamma_{i}=0$ ($i=c1,c3$), there is no solution except for the trivial one with $y_e = y_q=y_u=y_l=0$. The single pole structure only shows up trivially and preserves to higher loop orders. 
For an abelian gauge theory, $\gamma_g$ cannot be zero for non-vanishing $\gamma_i$ and the shifted-pole structure cannot be achieved at one loop level (or higher-order levels).  Note that if we instead consider non-abelian gauge theories, it is possible to have $\gamma_g = 0$ for the one-loop beta function and observe shifted-pole structure for the four-fermion celestial amplitudes.  
We further consider the possibility of $\gamma_{i}/\gamma_g$ being a negative integer. For simplicity, we require $y_q + y_u = 0$ to avoid operator mixing at one-loop level. We are able to find solution for $\gamma_{c3}/\gamma_g = -1$ ($y_e = -y_u, y_d = y_u, y_l = y_u$, $n_G=1$, $n_c =3$) for the running of $C^{(3)}_{lequ}$. This solution gives $\gamma_{c1}/\gamma_g = 3/5$. The celestial amplitude for $\mathcal{A}(L_\alpha Q_{a\beta} u_{\dot{a}}e)$ thus has both pole structure of order $2$ at $\beta = -2$ and branch cut structure for $\beta < -2$.  
With these hypercharges, however, the $U(1)_Y^3$ anomaly cancellation requires existences of heavy chiral fermions charged under the $U(1)_Y$ gauge group.
Other negative integer values for $\gamma_{i}/\gamma_g$ can not be achieved unless we choose smaller values of $n_c$, e.g. to be one. Larger $n_G$ increases $\gamma_g$ which results in even smaller $|\gamma_{i}/\gamma_g|< 1$, evading solutions of negative integer values of $\gamma_{i}/\gamma_g$.

We note here again that the multipole structures in the celestial amplitude is due to the termination of leading log contribution (to $C^{(3)}_{lequ}$) at finite loop order.  This feature should be accidental, and would not hold, for instance, 
once the resummation of sub-leading log terms or contributions from more than one dimensionless couplings are included. 
Due to the difficulties of solving the RGEs and performing the corresponding Mellin transformations, in general we are not able to do analytical calculations for higher loop-order contributions. 
However, we do note that in certain $SU(N)$ non-abelian gauge theories with coupling $g$, $\gamma_g \neq 0$ at the one-loop order, while with higher order loop contributions
the theory can flow to an interacting fixed point in the IR~\cite{Belavin:1974gu, PRL.33.244, Terning:2006bq, Aoki:2012ve} for certain numbers of fermion flavors charged. 
At this fixed point, the gauge coupling $g$ does not run, and the celestial amplitude generated by the higher dimensional operators have shifted-pole structures (as in \autoref{eq:mellinshift}) instead of branch cuts.


\section{Conclusions}
\label{sec:secv}

The analytic structure of celestial amplitudes in the negative $\beta$-plane encodes information on the EFT in the deep IR. 
Loop contributions generate $\log^n\omega$ terms in the amplitude, which naively map to multipoles structures at (negative) integer values of $\beta$.  However, as the log series diverges in the $\omega\to 0$ limit and needs to be resummed, so is the celestial amplitude in the region near the poles.  To obtain the correct analytic structure, one needs to first resum the log contributions in the amplitude before doing the Mellin transformation.  In this paper, we examine the analytic structure of celestial amplitudes for massless EFTs with this treatment, focusing on the leading log contributions.  Not surprisingly, the results differ significantly from the multipole structures of fixed-order loop contributions.  Branch cuts are ubiquitous, which originate from incomplete Gamma functions, generated by the Mellin transform of amplitudes of the form $\sim (1-\gamma \log\omega)^{-r}$.  
Different structures in the celestial amplitudes are also observed if the anomalous dimensions satisfy certain conditions.    
Instead of branch cuts, shifted-single pole can be generated in theories in which the dimensionless couplings do not run at leading order.  It is also possible to tune the anomalous dimensions in a 4-fermion amplitude to make the leading log contributions terminate at a certain order, in which case the single or multipole structures remain under the resummation of leading log terms.  These pole structures are likely to be accidental and unstable under slight changes in the anomalous dimensions, for instance with the inclusion of next-leading log contributions.

As we argued above, the resummation of the leading log contributions resolves the large log problem in the deep IR and generates meaningful celestial amplitudes in regions near the negative integers on the $\beta$-plane.  It is yet unclear whether the leading log contributions capture at least qualitatively all possible analytical structures in the negative $\beta$-plane, or if new structures can be generated by higher-order contributions.  To include the next-leading log contributions, two-loop RGEs are needed, which can be difficult to solve analytically.   Furthermore, a complicated function of $\omega$ can also be difficult to Mellin transform analytically.  
In massive theories, the $\omega$ dependences in the amplitudes are also generally more complicated.  The investigation of the analytic structures of celestial amplitudes in these more complicated cases turn out to be challenging, and may require novel tools and strategies. 

While the analytic structures of momentum space amplitudes have clear physical meanings for either poles or branch cuts, the implications of those in celestial amplitudes are far less clear.  It is also difficult to write down dispersion relations in the $\beta$-plane (the ones in Ref.~\cite{Chang:2021wvv} are in the $\omega$-plane), as the behavior of the contour at infinity is not well understood.  A better understanding of the physics information encoded in the celestial amplitudes may also tell us what analytic structures are expected in general.  


\section*{Acknowledgements}
We would like to thank Florian Herren, Hongliang Jiang and Ding Yu Shao for useful discussions. JG is supported by National Natural Science Foundation of China (NSFC) under grant No.~12035008. Fermilab is operated by Fermi Research Alliance, LLC under contract number DE-AC02-07CH11359 with the United States Department of Energy. LTW is supported by the DOE grant DE-SC0013642.


\appendix
\section{Celestial Sphere Kinematics}
\label{sec:celesphere}
Focusing on the massless amplitude, the null four-momenta $p_{\alpha \dot{\alpha}}$ can be written in terms of spinor-helicity variables, $p_{\alpha \dot{\alpha}} = \lambda_\alpha\tilde{\lambda}_{\dot{\alpha}}$. With specific choice of frame, it is natural to write
\begin{eqnarray}
\lambda_\alpha=\eta \sqrt{2\omega}
\begin{pmatrix}
1\\z
\end{pmatrix},
~~\tilde{\lambda}_{\dot{\alpha}}=\sqrt{2\omega}
\begin{pmatrix}
1\\\bar{z}
\end{pmatrix},
\end{eqnarray}
where $\eta =\pm$ corresponds to outgoing/incoming particles. The variable $z$ gives the direction of the null momentum and specifies a point on the celestial sphere as one can write 
\begin{equation}
p = \eta \omega  (1+ z \bar{z}, z+ \bar{z}, -i (z-\bar{z}), 1-z\bar{z}) = \eta \omega  (1+ |z|^2, \re(z), \im (z), 1-|z|^2), 
\end{equation}
where for the second equlity, we use the fact that for real momentum, $\bar{z} = z^{*}$. The inner product in terms of the helicity spinors are written as
\begin{eqnarray}
\left<ij\right> = \epsilon^{\alpha\beta}\lambda_{j,\alpha}\lambda_{i, \beta} = 2\eta_i\eta_j\sqrt{\omega_i\omega_j}z_{ij}, \left[ij\right] = \epsilon^{\dot{\alpha}\dot{\beta}}\tilde{\lambda}_{j,\dot{\alpha}}\dot{\lambda}_{i, \dot{\beta}} = 2\sqrt{\omega_i\omega_j}\bar{z}_{ij},
\end{eqnarray}
with $z_{ij} = z_i-z_j$ and $\bar{z}_{ij} = \bar{z}_i-\bar{z}_j$. The Mandelstam variable $s_{ij}$ satisfies the following relation;
\begin{eqnarray}
s_{ij}= \left<ij\right>\left[ij\right], ~~\frac{z_{ij}}{\bar{z}_{ij}} = \eta_i\eta_j \frac{\left<ij\right>^2}{s_{ij}}.
\end{eqnarray}
Consider the kinematic configuration with the incoming state in the $s-$ channel for massless 4-point amplitude $12\rightarrow 34$, we have $\eta_1 = \eta_2 = -\eta_3=-\eta_4$. In this case, to have momentum conservation satisfied, the factor $ 0 \leq z = \frac{z_{13}z_{24}}{z_{12}z_{34}}=\bar{z} \leq 1$. Setting $s = \omega^2$, we have $t = -z\omega^2 \leq 0~(13\rightarrow 24)$, $u = -(1-z)\omega^2\leq 0 ~(14\rightarrow 23)$. 
Notice that $z = \frac{1-\cos\theta}{2}$ with $\theta$ being the scattering angle in the center of mass frame, $z = 0$ corresponds to the forward limit.  
Using the spinor brackets, momentum conservation also gives the following equities:
\begin{eqnarray}
\frac{\left<23\right>}{\left<13\right>}  t + \frac{\left<24\right>}{\left<14\right>} u = 0, ~
\frac{\left<12\right>}{\left<32\right>}  u + \frac{\left<14\right>}{\left<34\right>} s = 0, ~ \frac{\left<12\right>}{\left<42\right>}  t + \frac{\left<13\right>}{\left<43\right>} s = 0. 
\end{eqnarray}
With the above relations, we can rewrite, e.g. $\left[13\right]\left[24\right]$ and $\left[14\right]\left[23\right]$  as:
\begin{eqnarray}
\left[13\right]\left[24\right] = t \bigg(\frac{\bar{z}_{12}\bar{z}_{34}}{z_{12}z_{34}}\bigg)^{1/2}, 
~~~\left[14\right]\left[23\right] = -u \bigg(\frac{\bar{z}_{12}\bar{z}_{34}}{z_{12}z_{34}}\bigg)^{1/2}.
\end{eqnarray}
The functions of $z_{ij} (\bar{z}_{ij})$ here quantifies the non-trivial action of the Lorentz group on the external massless states. Additional functions of $z_{ij} \bar{z}_{ij}$ are also required to match $\mathcal{A}(\beta, z)$ to the convential celetial amplitude, with more details shown in~\cite{Arkani-Hamed:2020gyp}.

\section{Properties of the incomplete $\Gamma$ function}
\label{sec:Gamma}
Here we provide more details on the incomplete Gamma function and the corresponding celestial amplitude.
Considering $r  = 1, 2,  3, ...$, $\Gamma(1-r, x)$ can be expanded as~\cite{wikiInGamma}:
\begin{eqnarray}
\Gamma(1-r, x) &=& \frac{1}{(r-1)!}\bigg( \frac{e^{-x}}{x^{r-1}}\sum^{r-2}_{k=0} (-1)^k(r-k-2)!x^k + (-1)^{r-1} \Gamma(0,x)\bigg),
\end{eqnarray}
with $\Gamma(0, x) = -{\rm EulerGamma} -\log(x) -\sum^{\infty}_{k=1}\frac{(-x)^k}{k(k)!}$. When $r=1$, the summation above equals zero. We thus have for $\mathcal{A} (\beta)$ in \eq{mellinoneloop}:
\begin{eqnarray}
\label{eq:im1}
&&\mathcal{A} (\beta) =\notag\\ 
&&c_{a_M}\frac{1}{\Gamma(r) }\frac{M^{\beta+a}}{\kappa} \bigg(\exp {\rho(\beta+a)} (\frac{-\beta-a}{\kappa})^{r - 1}\Gamma(0,\rho(\beta+a)) 
+ \sum^{r-2}_{k=0}(r-k-2)!\rho^k(-\beta-a)^k\bigg).\notag\\ 
\end{eqnarray}
In addition, the logarithmic function in $\Gamma(0, x)$ indicates branch cut when $\beta+ a $ crosses the negative real axis. The branch cut can be derived from the imaginary part of $\mathcal{A}(\beta)$ as:
\begin{eqnarray}
&&{\rm Disc}\mathcal{A} = 2 i\im\mathcal{A} (\beta )\bigg|_{\beta<-a} = - 2c_{a_M} i\frac{\pi}{\Gamma(r)}
\frac{M^{\beta+a}}{\kappa}\exp{\rho(\beta+a)}(\frac{\beta+a}{\kappa})^{r-1}.
\end{eqnarray} 
As $r = 1, 2, ...$ and $\Gamma(0, x)$ only diverges logarithmically as $\beta+ a\rightarrow 0$, there are no single-pole or multi-pole structure at $\beta  = -a$. It is only when $r = 1$ that the real part of the celestial amplitude diverges logarithmically at $\beta+ a =0$.

For $r\neq  1, 2, ...$, one could instead use the following expansion~\cite{DLMFInGamma} 
\begin{equation}
\label{eq:expansion}
\Gamma(1-r, x)=-x^{1-r} e^{-x} \sum^{\infty}_{k=0} \frac{x^k}{(1-r)(2-r)...(k+1-r)} + \Gamma(1-r),
\end{equation}
which reduces to $\Gamma(1-r,x) = (-r)!e^{-x} \underset{k=0}{\overset{-r}{\sum}} \frac{x^k}{k!}$ for $r$ being non-positive integer. One can identify that
\begin{eqnarray}
\label{eq:im2}
\mathcal{A} (\beta)&=& 
{\rm finite} +  c_{a_M}\frac{M^{\beta+a}}{\kappa} \exp{\rho(\beta+a)}\bigg(\frac{\beta+a}{\kappa}\bigg)^{r -1}\Gamma(1-r),
\end{eqnarray}
where the {\rm finite} piece comes from the fact that the absolute value of the summation over $k$ is smaller than $\alpha \times e^{|\beta+a| / \kappa}$ with $\alpha >0$. As $\Gamma(x)$ is finite for $x\neq 0, -1, -2, ...$, the only singularity possible in the above celestial amplitude is at $\beta = -a$ region.  
Notice that due to the $(\beta+a)^{r-1}$ term, the celestial amplitude also has branch cut for $\beta+a <0$ if $r-1$ is non-integer. We then have:
\begin{eqnarray}
\label{eq:im2}
{\rm Disc}\mathcal{A} = 2 i\im\mathcal{A} \bigg|_{\beta<-a} = -2c_{a_M}i
\frac{M^{\beta+a}}{\kappa}\exp{\rho(\beta+a)}\bigg(\frac{|\beta+a|}{\kappa}\bigg)^{r-1}\Gamma(1-r) \sin(\pi r).\notag\\ 
\end{eqnarray}
The above equations shows that the branch cut vanishes when $r$ is nonpositive interger $r \in Z^-$. Notice that when $r = 0$, the real part has single pole at $\beta = -a$ as expected.


\bibliographystyle{JHEP}
\bibliography{cele1}

\end{document}